\documentclass[preprintnumbers,article,amsmath,amssymb,floatfix,10pt,prd,superscriptaddress,nofootinbib,twocolumn]{revtex4}

\usepackage{doi}

\usepackage{graphicx}
\usepackage{dcolumn}
\usepackage{bm}
\usepackage{color}
\usepackage{enumitem}
\usepackage{amsmath}
\usepackage{amssymb}

\newcommand{\beq}{\begin{equation}}
\newcommand{\eeq}{\end{equation}}
\newcommand{\bea}{\begin{eqnarray}}
\newcommand{\eea}{\end{eqnarray}}

\usepackage{bbm}
\usepackage{amsfonts}
\usepackage{mathrsfs}
\usepackage{latexsym}
\usepackage{epsfig}
\usepackage{epstopdf}
\usepackage{epstopdf}
\usepackage{graphicx}
\usepackage{amssymb}
\usepackage{amsmath}
\usepackage{dcolumn}
\usepackage{bm}
\usepackage{color}
\usepackage{comment}
\usepackage{xcolor}
\usepackage{ulem}
\usepackage{hyperref}
\hypersetup{
  colorlinks=true,        
  linkcolor=magenta,         
  citecolor=red,         
}
\begin{document}

\title{Light Deflection and Greybody Bound Around a BTZ-ModMax Black Hole in Plasma Medium}



\author{Ritesh Pandey}
\email{riteshphy0@gmail.com}
\affiliation{Centre for Space Research, North-West University, Potchefstroom 2520, South Africa}

\author{Shubham Kala}
\email{shubhamkala871@gmail.com}
\affiliation{The Institute of Mathematical Sciences, C.I.T. Campus, Taramani, Chennai--600113, Tamil Nadu, India}

\author{Amare Abebe}
\email{Amare.Abebe@nithecs.ac.za}
\affiliation{Centre for Space Research, North-West University, Potchefstroom 2520, South Africa}
\affiliation{National Institute for Theoretical and Computational Sciences (NITheCS), South Africa}

\author{Hemwati Nandan}
\email{hnandan@associated.iucaa.in}
\affiliation{Department of Physics, Hemvati Nandan Bahuguna Garhwal Central University, Srinagar, Garhwal, Uttarakhand--246174, India}
\affiliation{Centre for Space Research, North-West University, Potchefstroom 2520, South Africa}

\author{G.~G.~L.~Nashed}
\email{nashed@bue.edu.eg}
\affiliation{Centre for Theoretical Physics, The British University in Egypt, Cairo 11837, Egypt}
\affiliation{Centre for Space Research, North-West University, Potchefstroom 2520, South Africa}

\begin{abstract}
We study the deflection of light in a homogeneous plasma medium around a BTZ–ModMax black hole, focusing on the effects of the ModMax nonlinear electrodynamics parameter and the cosmological constant. Using the Gauss–Bonnet theorem applied to the corresponding optical geometry in plasma, we derive a modified expression for the deflection angle and examine how plasma dispersion alters the gravitational lensing behavior. The influence of the ModMax parameter in the presence of homogeneous plasma is compared with its vacuum counterpart, as well as with the charged and static BTZ black hole cases, revealing distinct signatures arising from nonlinear electrodynamics. This work highlights the combined impact of homogeneous plasma, spacetime curvature, and nonlinear field dynamics on light deflection in lower-dimensional black hole geometries. We further study the greybody factor and analyze how the presence of homogeneous plasma and the ModMax parameter modifies the energy emission spectrum of the black hole. Our results demonstrate that both plasma effects and nonlinear electrodynamics significantly influence the transmission probabilities and emission rates, providing deeper insight into wave propagation and observational signatures in lower-dimensional black hole geometries.
\end{abstract}

\maketitle
\section{Introduction}

One of the most fascinating phenomenon predicted by general relativity are black holes. They provide a natural environment to examine the behavior of matter, radiation, and spacetime under conditions of severe gravity~\cite{einstein1916foundations,chandrasekhar1998mathematical,Bardeen:1970zz}. Astrophysical black holes (BH) are intrinsically higher-dimensional phenomena; yet, lower-dimensional models—particularly black holes in (2+1)-dimensional spacetime—have proven valuable in clarifying basic concepts of gravitational physics~\cite{witten1989topology}. Such simplified models retain key characteristics of real BH, such as event horizons, thermodynamics properties and causal structure, allowing for a highly organized analytical study~\cite{Cardenas2014}. Thus, three-dimensional black holes are also important theoretical models in terms of understanding the effects of exotic matter fields, modified electrodynamics and both classical and quantum aspects of gravity. Especially in regions with high electromagnetic fields located close to the event horizon, black hole solutions in ModMax electrodynamics exhibit substantial deviations from their Maxwell counterparts ~\cite{kruglov2021generalized,kuzenko2021duality,avetisyan2021democratic}. These deviations may have significant impacts on the nature of particles and photons and on the structure of spacetime. For the case of (2+1)-dimensional gravity, the coupling of ModMax electrodynamics to the Baños-Teitelboim-Zanelli (BTZ) black hole creates a robust background for the study of consequences for the geometry and physical properties of BH from nonlinear electromagnetic processes~\cite{Cardenas2014,Hendi2018,Nashed2018,Mu2020,Canate2020,EslamPanah2023a,Karakasis2023,EslamPanah2023b}. These models are particularly significant for examining the impact of strong-field electrodynamics on observable phenomena, including gravitational lensing and photon trajectories~\cite{Jaki1978,Darwin1959}. The BTZ black hole is the most remarkable lower-dimensional solution owing to its solvable nature and asymptotically anti-de Sitter (AdS) geometry~\cite{Quan:2026awx}. Since the first discovery of the BTZ metric, there have been numerous extensions of it to incorporate a variety of physical features such as rotation, electric charge, nonlinear electromagnetic sources, scale-dependent coupling constants, and modifications in the gravity sector~\cite{Upadhyay:2023yhk,kala2025propagation}. All of these generalizations revealed an intricate physics involving complicated horizons structure, changed parameters for thermodynamics and new optical features. Recently, much interest has been directed on black holes exhibiting non-linear electrodynamics ~\cite{EslamPanah:2024lbk,kala2025propagation,Javed:2025wty}. This is aimed at regularizing singularities, probing strong electromagnetic fields and detecting deviations from the Maxwell field theory. The Modified Maxwell (ModMax) theory exemplifies non-linear electrodynamics. It preserves conformal invariance and electric-magnetic duality while including nonlinear alterations to the standard Maxwell Lagrangian~\cite{de2014holography}. BH solutions derived from ModMax electrodynamics show considerable divergences from their linear equivalents, particularly in high-field regions next to the event horizon. In three-dimensional gravity, ModMax-corrected BTZ black holes serve as a vital foundation for investigating the impact of nonlinear electromagnetic processes on spacetime geometry, photon dynamics, and observable effects, including gravitational lensing~\cite{kala2025propagation}. Gravitational lensing, an effect caused by spacetime curvature leading to bending of light, is among the most efficient ways to study black hole structures and gravitational fields~\cite{Iyer2007,Cunha2018,Mustafa:2022xod,Atamurotov:2023rye,Kala2020b,Kala2022,Kala2024,pandey2026geodesics}. It has been instrumental in studying and restricting the scope of alternative gravitational theories, especially since the confirmation of Einstein's predictions made during the famous 1919 solar eclipse. In the strong-gravity domain, light propagation near BH produces complex phenomena, such as significant deflection angles, relativistic images, photon spheres, and optical caustics~\cite{iyer2007light}. These effects are highly sensitive to the geometry of the underlying space and medium, thereby making gravitational lensing a superior method of studying modified black hole space-times.
In addition to vacuum propagation, in the real astrophysical environment, one can find some dispersive medium, such as plasma as discussed in ~\cite{Broderick2003,Bicak1975,Kichenassamy1985,Krikorian1999,Synge1960,Perlick:2015vta,Javed:2021ymu,Babar:2021nst}. Plasma affects photon propagation because it changes the path of photons by creating an effective refraction index that deflects light according to its frequency. Such effect could have a significant impact on lensing parameters, particularly when close to some compact object where the density of plasma is greater. Hence, studying plasma gravitational lensing is quite essential for gaining knowledge about BH surroundings and making more precise predictions~\cite{Perlick:2017fio,Upadhyay:2023yhk,Kala:2025fld,Roy:2026kzf,Babar:2021exh,Pahlavon:2024caj}. Recently, several analytical techniques have been devised to investigate light deflection in curved spacetime. The Gauss–Bonnet theorem has been shown to be an effective and sophisticated method for determining weak deflection angles in optical geometry, supplementing the traditional geodesic-based approach~\cite{gibbons2008applications,Ovgun:2018fnk}. This geometric methodology provides a coordinate-invariant framework and has been effectively used in several black hole spacetimes, including those characterized by spin, charge, cosmological constants, and surrounding substances such as plasma. This study examines gravitational lensing induced by a ModMax BTZ black hole inside a plasma medium, prompted by recent advancements. Our study examines the aggregate impact of nonlinear electromagnetic phenomena and plasma-induced dispersion on photon trajectories and light deflection within a lower-dimensional AdS framework. We calculate the weak deflection angle using the Gauss–Bonnet theorem and analyze the impact of key variables, including the ModMax coupling, black hole charge, plasma frequency, and cosmological constant, by formulating the optical metric and evaluating the Gaussian curvature. This work offers novel insights into the optical characteristics of black holes influenced by nonlinear electrodynamics, emphasizing the synergistic effects of altered electromagnetism and plasma conditions on gravitational lensing in (2+1)-dimensional spacetimes. Our findings provide a substantial theoretical foundation for investigating the effects of exotic matter and may facilitate the development of more accurate models of light propagation in intense gravitational fields.\\
Hawking Radiation represents a basic quantum property of BH. The thermal radiation produced by BH as a result of quantum effects on the event horizon is known as Hawking Radiation~\cite{page1976particle}. Practically, the spacetime geometry in the vicinity of black holes causes changes in the emission of BH. Thus, the radiation does not behave like a blackbody anymore. The changes are caused by the presence of Greybody factors. According to the analysis of Greybody factors, we are able to predict whether the particles which emerge after passing the effective potential barrier may escape infinity. The factors also depend on the mass, charge, and angular momentum of black holes along with the type of surrounding fields ~\cite{Boonserm2008,Dai2010,Fernando2005,Mistry2017,Kanti2002,Kanti2005,Gogoi:2023fow}. They are crucial for determining observable phenomena, such as the energy emitted by black holes and the energy they can assimilate. In this work, greybody variables for a charged BTZ–ModMax black hole in a plasma environment are examined.  In the recent time greybody factor and lensing has been studied~\cite{javed2022effect}. These results are essential for understanding the dissemination of radiation in authentic astrophysical environments and provide improved insight into the observational characteristics linked to lower-dimensional black hole models. 

The paper is organized as follows. In Sec.~\ref{sec2}, we discuss the spacetime structure and horizon properties of the BTZ--ModMax black hole. In Sec.~\ref{sec3}, we study the effects of a homogeneous plasma, including the equations of motion, effective potential, photon sphere, and the distance of closest approach. Sec.~\ref{sec4} is devoted to the gravitational lensing of the BTZ--ModMax black hole in a plasma medium. In Sec.~\ref{sec5}, we analyze the bounds on the greybody factor in the presence of plasma. Finally, in Sec.~\ref{sec6}, we summarize our results and present the conclusions. Throughout this work, we employ geometric units ($G = c = 1$), in which all physical quantities can be expressed in terms of a single fundamental unit (typically length). The ModMax parameter $\gamma$ is dimensionless, while the cosmological constant $\Lambda$ has dimensions of inverse length squared.


\section{BTZ-ModMax Black Hole Spacetime} \label{sec2}
The action delineating Einstein gravity coupled with ModMax nonlinear electrodynamics (NLED) in the context of a cosmic constant inside a three-dimensional spacetime is expressed as ~\cite{Bandos:2020jsw}
\begin{equation}
I = \frac{1}{16\pi} \int_{\partial M} d^3x \, \sqrt{-g} \left[ R - 2\Lambda - 4 \mathcal{L} \right],
\end{equation}
where $R$ denotes the Ricci scalar, $\Lambda$ is the cosmological constant, and $g = \det(g_{\mu\nu})$ represents the determinant of the metric tensor $g_{\mu\nu}$. The quantity $\mathcal{L}$ corresponds to the ModMax Lagrangian~\cite{Kosyakov:2020wxv}.

In this work, we assume that the ModMax Lagrangian in three dimensions retains a structure analogous to its four-dimensional counterpart, and is given by
\begin{equation}
\mathcal{L} = X \cosh \gamma - \sqrt{X^2 + Y^2} \sinh \gamma,
\end{equation}
where $\gamma$ is a dimensionless parameter that characterizes the strength of the nonlinear electrodynamics. The quantities $X$ and $Y$ are Lorentz invariants, with $X$ being a scalar and $Y$ a pseudoscalar, defined as
\begin{equation}
X = \frac{F}{4}, \qquad Y = \frac{F_e}{4},
\end{equation}
where $F = F_{\mu\nu} F^{\mu\nu}$ is the Maxwell invariant. The electromagnetic field tensor $F_{\mu\nu}$ is defined as
\begin{equation}
F_{\mu\nu} = \partial_{\mu} A_{\nu} - \partial_{\nu} A_{\mu},
\end{equation}
where $A_{\mu}$ is the gauge potential. Furthermore, $F_e = F_{\mu\nu} \tilde{F}^{\mu\nu}$, where the dual tensor is given by
\begin{equation}
\tilde{F}_{\mu\nu} = \frac{1}{2} \epsilon_{\rho\lambda\mu\nu} F^{\rho\lambda}.
\end{equation}
It is worth noting that in the limit $\gamma = 0$, the ModMax Lagrangian reduces to the standard linear Maxwell form, i.e., $\mathcal{L} = \frac{F}{4}$.

\noindent The corresponding BTZ black hole solution in the framework of ModMax nonlinear electrodynamics is therefore given by~\cite{panah2024thermodynamics}
\begin{equation} \label{metric}
ds^{2} = -\psi(r)\, dt^{2} + \frac{dr^{2}}{\psi(r)} + r^{2} d\phi^{2},
\end{equation}
where the lapse function takes the form
\begin{equation} \label{mfunction}
\psi(r) = -m_{0} - \Lambda r^{2} - 2 q^{2} e^{-\gamma} 
\ln\!\left(\frac{r}{l}\right).
\end{equation}
Here, $m_{0}$ is an integration constant associated with the ADM mass of the black hole, $q$ denotes the electric charge, and $l$ is an arbitrary length scale. The constant $\Lambda$ represents the cosmological constant, which we take to be negative throughout this work. The parameter $\gamma$ characterizes the strength of the ModMax nonlinear electrodynamics; when $\gamma = 0$, the metric reduces to the usual charged BTZ black hole in the Einstein–$\Lambda$–Maxwell theory~\cite{Kosyakov:2020wxv}.

\subsection{Horizons of the BTZ--ModMax Black Hole}

The horizons of the BTZ--ModMax black hole are determined by the real, positive roots of the lapse function defined in  Eq.~(\ref{mfunction}).

Setting $\psi(r_h)=0$, one obtains the transcendental equation
\begin{equation}
-m_0 - \Lambda r_h^2 - 2 q^2 e^{-\gamma}
\ln\!\left(\frac{r_{h}}{l}\right)=0,
\end{equation}
which cannot be solved algebraically due to the logarithmic contribution originating from the ModMax nonlinear electrodynamics.
The horizon radius can be expressed in terms of the Lambert $W$ function and is given by~\cite{Hendi:2010px,kala2025propagation},
\begin{equation}
r_+ = l \exp\!\left[
-\frac{1}{2}
W_{-1}\!\left(
-\frac{\Lambda r_0^2}{q^2 e^{-\gamma}}
\exp\!\left(-\frac{m_0}{q^2 e^{-\gamma}}\right)
\right)
+\frac{m_0}{2 q^2 e^{-\gamma}}
\right],
\end{equation}
\begin{equation}
r_- = l \exp\!\left[
-\frac{1}{2}
W_{0}\!\left(
-\frac{\Lambda r_0^2}{q^2 e^{-\gamma}}
\exp\!\left(-\frac{m_0}{q^2 e^{-\gamma}}\right)
\right)
+\frac{m_0}{2 q^2 e^{-\gamma}}
\right],
\end{equation}
where $W_0$ and $W_{-1}$ denote the principal and lower real branches of the Lambert $W$ function, respectively.
These branches are defined as~\cite{corless1996lambert}
\begin{equation}
W_0:\left[-\frac{1}{e},\infty\right)\rightarrow[-1,\infty),
\hspace{0.2cm}
W_{-1}:\left[-\frac{1}{e},0\right)\rightarrow(-\infty,-1],
\end{equation}
with the common value
\begin{equation}
W_0\!\left(-\frac{1}{e}\right)
=
W_{-1}\!\left(-\frac{1}{e}\right)
=
-1.
\end{equation}
The extremal BTZ--ModMax black hole configuration is obtained when the argument of the Lambert $W$ function reaches its critical value,
\begin{equation}
-\frac{\Lambda l^2}{q^2 e^{-\gamma}}
\exp\!\left(-\frac{m_0}{q^2 e^{-\gamma}}\right)
= -\frac{1}{e},
\end{equation}
for which the inner and outer horizons coincide, $r_+=r_-$.

The ModMax parameter $\gamma$ effectively rescales the electric charge contribution through the factor $q^2 e^{-\gamma}$, leading to significant modifications in the horizon structure and extremality conditions compared to the standard charged BTZ black hole.

\section{Effect of Homogenous Plasma } \label{sec3}

In this section, we analyse the effect of homogenoues plasma on the curvature of spacetime. We assume that spacetime time is filled with homogeneous plasma whose the electron plasma frequency $\omega_e$ is given by 
\begin{equation}
\omega_e^2(r) = \frac{4\pi e^2}{m}\, N(r),
\end{equation}
 where $N(r)$ is the number density of electrons in the plasma, $e$ and $m$ denote the charge and mass of the electron respectively. The refractive index satisfies the following relation \cite{perlick2015influence,Kala:2026bmz}
\begin{equation}
n^2(r) = 1 - \frac{\omega_e^2}{\omega_\infty^2} f(r),
\end{equation}
where $\omega_\infty$ is the light ray frequency detected by a static observer at infinity, while $\omega_e$ is the electron plasma frequency.
For further calculations, we consider the dimensionless ratio
\begin{equation}
\omega_p^2 = \frac{\omega_e^2}{\omega_\infty^2}.
\end{equation}

In the presence of a homogeneous plasma, the photon dispersion relation is given by
\begin{equation}
g^{\mu\nu} p_{\mu} p_{\nu} = -\omega_p^2,
\label{eq:dispersion}
\end{equation}
where $\omega_p$ is the constant plasma frequency. 
This relation replaces the vacuum null condition $g^{\mu\nu}p_{\mu}p_{\nu}=0$.

The corresponding Hamiltonian reads
\begin{equation}
\mathcal{H} = \frac{1}{2}\left(g^{\mu\nu}p_{\mu}p_{\nu} + \omega_p^2\right)=0.
\label{eq:hamiltonian}
\end{equation}

\subsection{Equations of Motion in Plasma}

We consider the BTZ ModMax metric
\begin{equation}
ds^2 = -\psi(r)dt^2 + \frac{dr^2}{\psi(r)} + r^2 d\phi^2,
\end{equation}
with the metric function defined in  Eq.~(\ref{mfunction}).

Due to the spacetime symmetries, the conserved quantities are
\begin{equation}
p_t = -E, \qquad p_{\phi} = L,
\end{equation}
where $E$ and $L$ denote the photon energy and angular momentum, respectively.

Using Eqs.~(\ref{eq:dispersion}) and (\ref{eq:hamiltonian}), the radial equation of motion becomes
\begin{equation}
\dot r^2 = E^2 - \psi(r)\left(\frac{L^2}{r^2} + \omega_p^2\right).
\label{eq:radial_plasma}
\end{equation}

\subsection{Effective Potential in Homogeneous Plasma}

From Eq.~(\ref{eq:radial_plasma}), the effective potential for photon motion in plasma can be identified as
\begin{equation}
V_{\text{eff}}(r)=\psi(r)\left(\frac{L^2}{r^2} + \omega_p^2\right),
\label{eq:veff_plasma}
\end{equation}
or explicitly,

\begin{equation}
V_{\text{eff}}(r)
=
\left[-m_0 - \Lambda r^2 - 2 q^2 e^{-\gamma} \\ln\!\left(\frac{r}{l}\right)\right]
\left(\frac{L^2}{r^2} + \omega_p^2\right)
\end{equation}
\begin{figure}[ht]
  \centering
  \includegraphics[width=85mm,height=75mm]{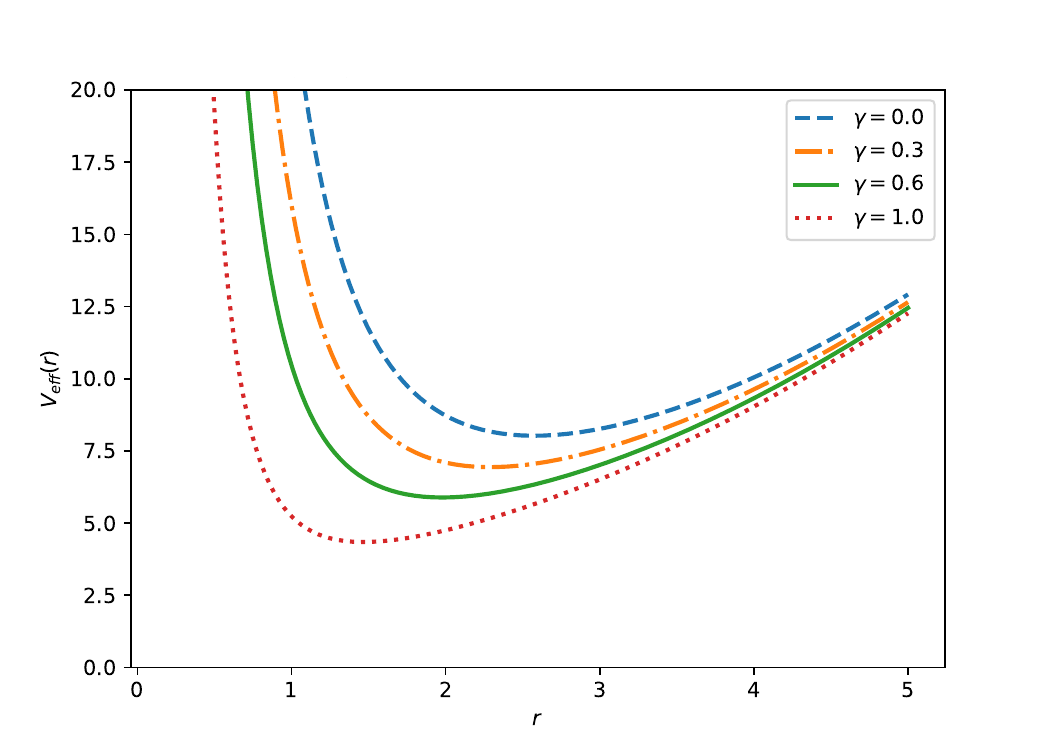}
\caption{Behaviour of effective potential with radial distance for different values of ModMax parameter in presence of plasma and fixed $m_0$ = 1, $l$ = 10, $q$ = 0.8, $\Lambda$ = 0.45, $L$ = 3.0, $\omega_{p}^{2}=0.5$ }

  \label{fig:V1}
\end{figure}
\begin{figure}[ht]
  \centering
  \includegraphics[width=85mm,height=75mm]{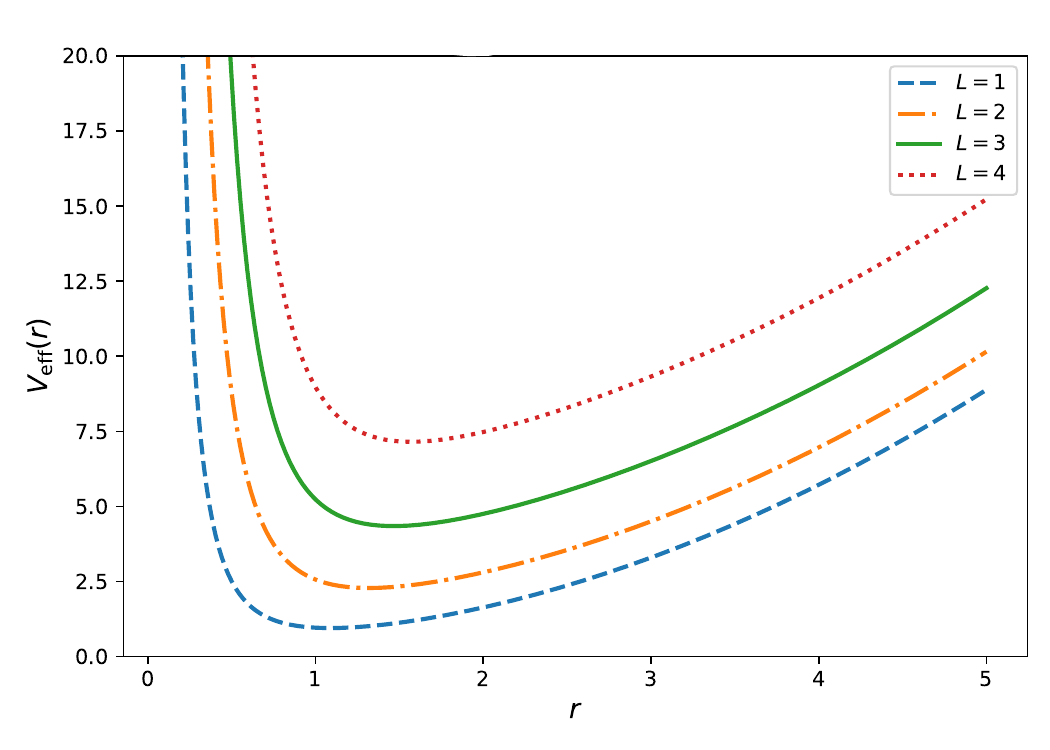}
\caption{Behaviour of effective potential with radial distance for different values of  angular momentum in presence of plasma and fixed  $m_0$ = 1, $l$ = 10, $q$ = 0.8, $\gamma$ = 1,  $\omega_{p}^{2}=0.5$}

  \label{fig:V2}
\end{figure}
In Fig.~\ref{fig:V1}, the effective potential is shown as a function of radial distance $r$ for different values of the ModMax parameter $\gamma$. It is observed that increasing $\gamma$ lowers the potential barrier, indicating a suppression of charge-induced effects due to nonlinear electrodynamics. In Fig.~\ref{fig:V2}, the effective potential is plotted for different values of angular momentum $l$. The height of the potential barrier increases with $l$, demonstrating that higher angular momentum leads to stronger centrifugal effects, thereby enhancing the stability of photon orbits.


\subsection{Photon Orbit in Plasma Environment}

In this section, we derive the photon circular orbit (photon sphere) in the presence of a homogeneous plasma background. 
The radius of the photon orbit $r_{\text{ph}}$ is determined by the condition for unstable circular null geodesics \cite{Kala2020b}.
\begin{equation}
\frac{dV_{\text{eff}}}{dr}\bigg|_{r=r_{\text{ph}}} = 0.
\end{equation}
Using Eq.~\eqref{eq:veff_plasma}, we obtain
\begin{equation}
\psi'(r)\left(\frac{L^2}{r^2} + \omega_p^2\right)
= \frac{2L^2}{r^3}\psi(r).
\label{eq:photon_condition_general}
\end{equation}
For the spacetime under consideration, the metric function is
defined in  Eq.~(\ref{mfunction}).
which yields
\begin{equation}
\psi'(r) = -2\Lambda r - \frac{2 q^2 e^{-\gamma}}{r}.
\end{equation}
Substituting these expressions into Eq.~\eqref{eq:photon_condition_general}, we obtain the equation governing the photon orbit:
\begin{align}
\left(-2\Lambda r + \frac{2 q^2 e^{-\gamma}}{r}\right)
\left(\frac{L^2}{r^2} + \omega_p^2\right) \\
= \frac{2L^2}{r^3}\left[-m_0 - \Lambda r^2 - 2 q^2 e^{-\gamma} \ln\left(\frac{r}{l}\right)\right].
\label{eq:photon_orbit_full}
\end{align}
Equation~\eqref{eq:photon_orbit_full} is a transcendental equation due to the logarithmic term and the presence of the plasma frequency. Therefore, the photon sphere radius $r_{\text{ph}}$ must, in general, be obtained numerically.

In the absence of plasma ($\omega_p = 0$), Eq.~\eqref{eq:photon_orbit_full} simplifies considerably. After straightforward algebraic manipulation, we obtain an analytical expression for the photon orbit:
\begin{equation}
r_{\text{ph}} = l \exp\left(-\frac{m_0 + q^2 e^{-\gamma}}{2 q^2 e^{-\gamma}}\right).
\end{equation}
This expression highlights the role of the nonlinear electrodynamics parameter $\gamma$ and charge $q$ in determining the photon sphere.
\begin{table}[t]
\centering
\caption{Numerically computed values of the photon orbit radius $r_{\mathrm{ph}}$ for different values of the ModMax parameter $\gamma$ and charge parameter $q$. The remaining parameters are fixed at $m_0 = 0.8$, $\Lambda = 1$, $l = 1.0$, $L = 1.0$, and homogeneous plasma parameter $\omega_p = 0.3$.}
\label{tab:rph_gamma_q}
\begin{tabular}{ccccc}
\hline
$\gamma \backslash q$ & 0.20 & 0.40 & 0.60 & 0.80 \\
\hline
0.10 & 1.7232 & 1.7117 & 1.6888 & 1.6425 \\
0.30 & 1.7238 & 1.7146 & 1.6969 & 1.6645 \\
0.50 & 1.7243 & 1.7170 & 1.7031 & 1.6793 \\
0.70 & 1.7248 & 1.7188 & 1.7078 & 1.6899 \\
1.00 & 1.7253 & 1.7209 & 1.7131 & 1.7010 \\
\hline
\end{tabular}
\end{table}

Table~\ref{tab:rph_gamma_q} shows the variation of the photon orbit radius $r_{\mathrm{ph}}$ with the ModMax parameter $\gamma$ and charge parameter $q$. It is observed that, for a fixed value of $\gamma$, the photon orbit radius decreases with increasing charge $q$. On the other hand, for a fixed charge, $r_{\mathrm{ph}}$ increases slightly with increasing $\gamma$. This indicates that the charge parameter introduces an effective attractive contribution, leading to an inward shift of the photon sphere, while the nonlinear electrodynamics effect encoded in $\gamma$ tends to counteract this behavior and produces a mild outward shift. In addition, the presence of plasma modifies the photon motion through the additional term $\omega_p^2$ in the effective potential. This introduces a frequency-dependent refractive index, which effectively alters the location of the photon sphere. In general, the plasma environment tends to shift the photon orbit outward and introduces deviations from the vacuum case, particularly in the low-frequency regime. These modifications can have observable consequences in black hole shadow and gravitational lensing studies.

\subsection{Distance of Closest Approach and Light Deflection in the BTZ--ModMax Spacetime}

The curvature of spacetime causes a photon to move in a curved path when it moves through the gravitational field of a small object. As the photon advances closer to the black hole, it expands to a minimum radial distance before being sent back out into space. The distance of closest approach  $r_0$ is the minimum distance at which the photon changes direction and the radial component of its velocity disappears i.e. $\dot{r}=0$.

For null geodesics in the BTZ--ModMax spacetime, the distance of closest approach is directly related to the impact parameter $b$ through
\begin{equation}
\frac{1}{b^2}
=
\frac{\psi(r_0)}{r_0^2},
\label{impact}
\end{equation}
where $\psi(r)$ denotes the lapse function of the BTZ--ModMax black hole and the impact parameter is defined as $b=L/E$, with $L$ and $E$ being the conserved angular momentum and energy of the photon, respectively. Eq.~\eqref{impact} follows from the null geodesic condition and encodes the influence of the spacetime geometry on photon propagation.
The ModMax parameter $\gamma$ alters the electromagnetic sector by using the exponential factor $e^{-\gamma}$, resulting in a diminished electric field contribution as $\gamma$ increases. Consequently, increased values of $\gamma$ extend the distance of closest approach to greater radii. Nonlinear electrodynamic phenomena result in a reduction in curvature. 

Using the relation in Eq.~\eqref{impact}, the total deflection angle of light can be written in an exact integral form \cite{iyer2007light,Kukreti:2025rzn}
\begin{equation}
\alpha(r_0)
=
2 \int_{r_0}^{\infty}
\frac{dr}{r}
\left[
\frac{r^2}{r_0^2}
\frac{\psi(r_0)}{\psi(r)}
-
1
\right]^{-1/2}
-
\pi .
\label{deflection_integral}
\end{equation}
This expression provides a formal description of light bending in the BTZ--ModMax spacetime and explicitly incorporates the effects of the cosmological constant, electric charge, and nonlinear electrodynamic parameter $\gamma$. In general, the integral in Eq.~\eqref{deflection_integral} does not possess a closed-form solution and must be assessed numerically. 

In the present work, however, we focus on the weak deflection regime, where photons propagate far from the black hole and the bending angle remains small. Rather than directly evaluating the integral in Eq.~\eqref{deflection_integral}, we employ the Gauss--Bonnet theorem to compute the weak-field deflection angle. This geometric method connects the deflection angle to the overarching characteristics of the optical metric, offering a robust and refined framework for examining gravitational lensing in lower-dimensional spacetimes. The details of this method and its application to the BTZ--ModMax black hole are presented in the following section.

\section{Gravitational Lensing of BTZ--ModMax Black Hole in Plasma Medium} \label{sec4}

This section analyzes the gravitational lensing of a BTZ black hole linked to ModMax nonlinear electrodynamics within the framework of a cold, non-magnetized plasma characterized by its refractive index $n(r)$.
The refractive index satisfies\cite{perlick2015influence},
\begin{equation}
n^2(r)=1-\frac{\omega_e^2}{\omega_\infty^2} f(r),
\end{equation}
where $\omega_\infty$ is the photon frequency measured by a static
observer at infinity and $\omega_e$ is the electron plasma frequency.

This approximation holds true when the spatial variation of electron density is insignificant relative to the scale of the photon trajectory,
\begin{equation}
\frac{1}{n_e}\frac{dn_e}{dr} \ll \frac{1}{b},
\end{equation}
where $b$ is the impact parameter. We further assume that the frequency of the photon is much higher than the frequency of the plasma, which is 
\begin{equation}
\frac{\omega_e^2}{\omega_\infty^2} \ll 1,
\end{equation}
which makes sure that dispersive effects stay miniature and the geometric optics approximation is still true. In these conditions, the refractive index is contingent solely on the underlying spacetime geometry, rather than on the spatial fluctuations of the plasma density. 

This approximation is particularly appropriate in the weak-field region ($r \gg r_h$), where the gravitational field is not very strong and realistic astrophysical plasma distributions can be considered as mostly uniform. But the homogeneous plasma approximation doesn't work near the black hole horizon or in places with strong density gradients, where the plasma frequency depends on where you are. 

Furthermore, the deflection angle is calculated using the Gauss--Bonnet theorem within the weak-field regime. This particularly requires the following conditions to hold:
\begin{equation}
\frac{m_0}{r} \ll 1, \quad \frac{q^2}{r^2} \ll 1, \quad m^2 c c_1 r \ll 1,
\end{equation}
ensuring that higher-order contributions such as $\mathcal{O}(q^4, m^4)$ can be neglected. In addition, the impact parameter must satisfy
\begin{equation}
b \gg r_h,
\end{equation}
to ensure that the light ray travels a long distance from the strong-field area and follows a trajectory that is approximately straight\cite{perlick2015influence}, 
\begin{equation}
r(\phi) = \frac{b}{\sin\phi}.
\end{equation}

Under these assumptions, the deflection angle remains small ($\tilde{\delta} \ll 1$), and the Gauss--Bonnet method yields the leading-order contribution to gravitational lensing. It is important to stress that this method does not take into account higher-order relativistic corrections or strong lensing effects close to the photon sphere, which is why it is different from exact methods that use null geodesic integration.

For the BTZ--ModMax black hole, the metric function is
\begin{equation}
f(r)=\Lambda' r^2 - m_0
- 2 q^2 e^{-\gamma}\ln\!\left(\frac{r}{l}\right),
\end{equation}
Here, we define $\Lambda' = -\Lambda$, and hence the refractive index becomes
\begin{equation}
n^2(r)=1-\frac{\omega_e^2}{\omega_\infty^2}
\left(\Lambda' r^2 - m_0
- 2 q^2 e^{-\gamma}\ln\frac{r}{l}\right).
\end{equation}
The corresponding optical metric in plasma is given by \cite{gibbons2008applications}
\begin{equation}
dt^2=g^{\text{opt}}_{ij}dx^idx^j
=\frac{n^2(r)}{f(r)^2}dr^2
+\frac{n^2(r)r^2}{f(r)}d\phi^2 .
\end{equation}
The determinant of the optical metric tensor is
\begin{equation}
\det g^{\text{opt}}
=\frac{n^4(r)r^2}{f^3(r)} .
\end{equation}
The Gaussian optical curvature is obtained from \cite{gibbons2008applications,Upadhyay:2023yhk}
\begin{equation}
K=
\frac{1}{\sqrt{g^{\text{opt}}}}
\left[
\partial_\phi(\sqrt{g^{\text{opt}}}\Gamma^\phi_{rr})
-\partial_r(\sqrt{g^{\text{opt}}}\Gamma^\phi_{r\phi})
\right].
\end{equation}
\begin{figure}[ht]
  \centering
  \includegraphics[width=85mm,height=75mm]{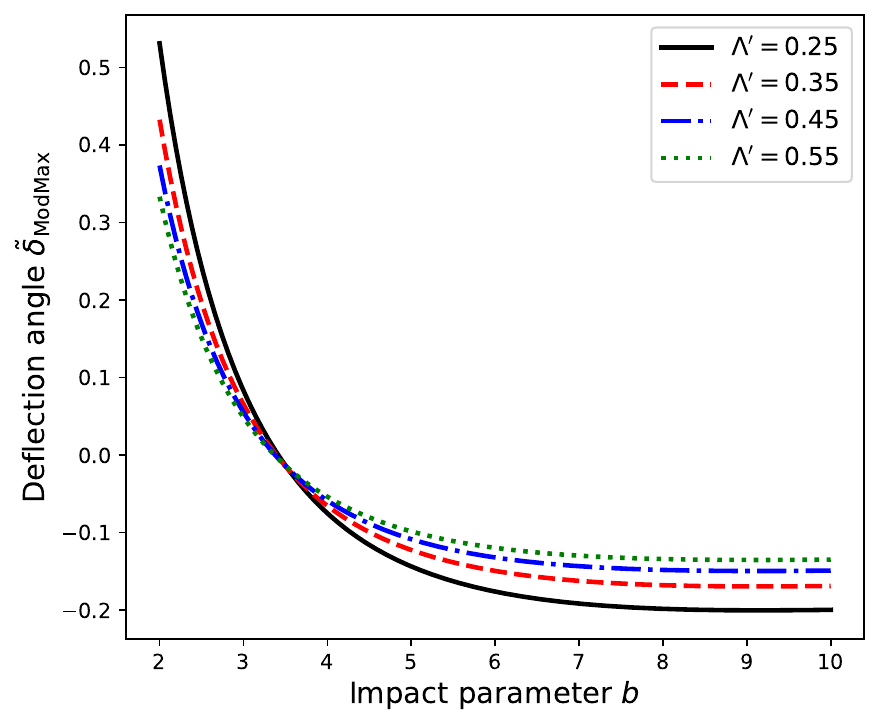}
\caption{Deflection angle behavior as a function of the impact parameter $b$ for various values of $\Lambda'$. The other parameters are fixed to $m_0 = 1$, $l = 10$, $q = 0.5$, $\gamma = 0.5$, and $\omega_p^2 = 0.5$.}
  \label{fig:da1}
\end{figure}

\begin{figure}[ht]
  \centering
  \includegraphics[width=85mm,height=75mm]{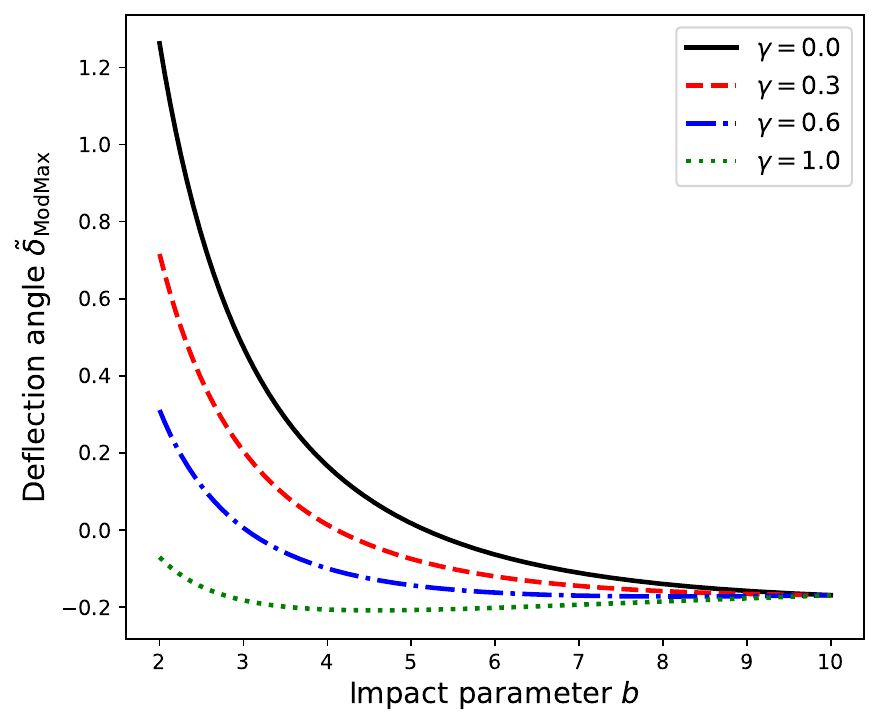}
\caption{Deflection angle behavior as a function of the impact parameter $b$ for various values of $\gamma$. The other parameters are fixed to $m_0 = 1$, $l = 10$, $q = 0.5$, $\Lambda' = 0.45$, and $\omega_p^2 = 0.5$.}

  \label{fig:da2}
\end{figure}
Since the optical metric is independent of $\phi$, this reduces to

\begin{equation}
K=-\frac{1}{\sqrt{g^{\text{opt}}}}
\partial_r\!\left(\sqrt{g^{\text{opt}}}\Gamma^\phi_{r\phi}\right).
\end{equation}
\begin{figure}[ht]
  \centering
  \includegraphics[width=85mm,height=75mm]{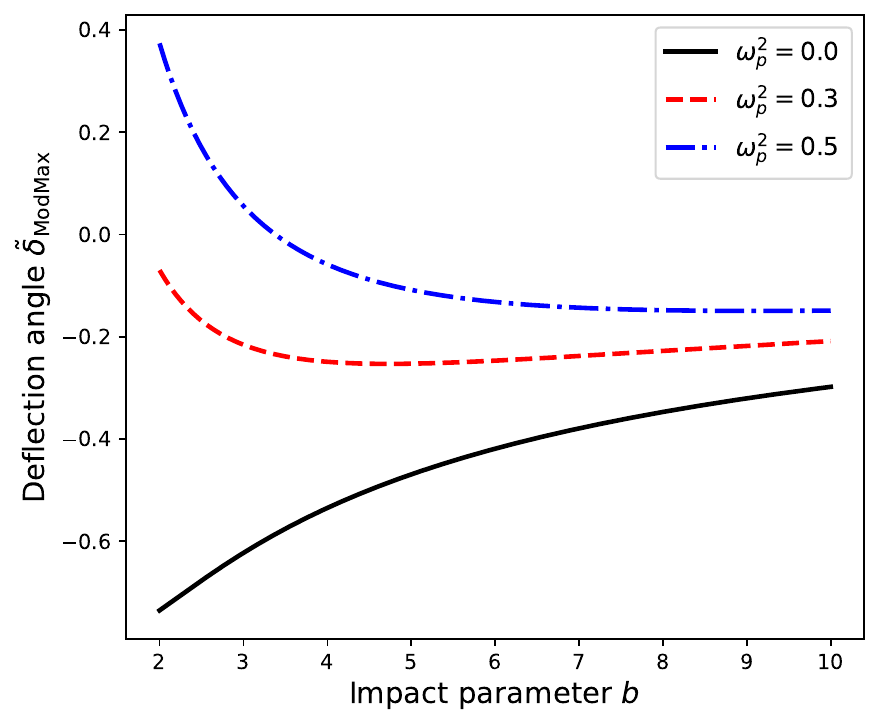}
\caption{Deflection angle behavior as a function of the impact parameter $b$ for various values of $\omega_p^2$.  The other parameters are fixed to $m_0 = 1$, $l = 10$, $q = 0.5$, $\Lambda' = 0.45$, and $\gamma = 0.5$. }

  \label{fig:da3}
\end{figure}

\begin{figure}[ht]
  \centering
  \includegraphics[width=85mm,height=75mm]{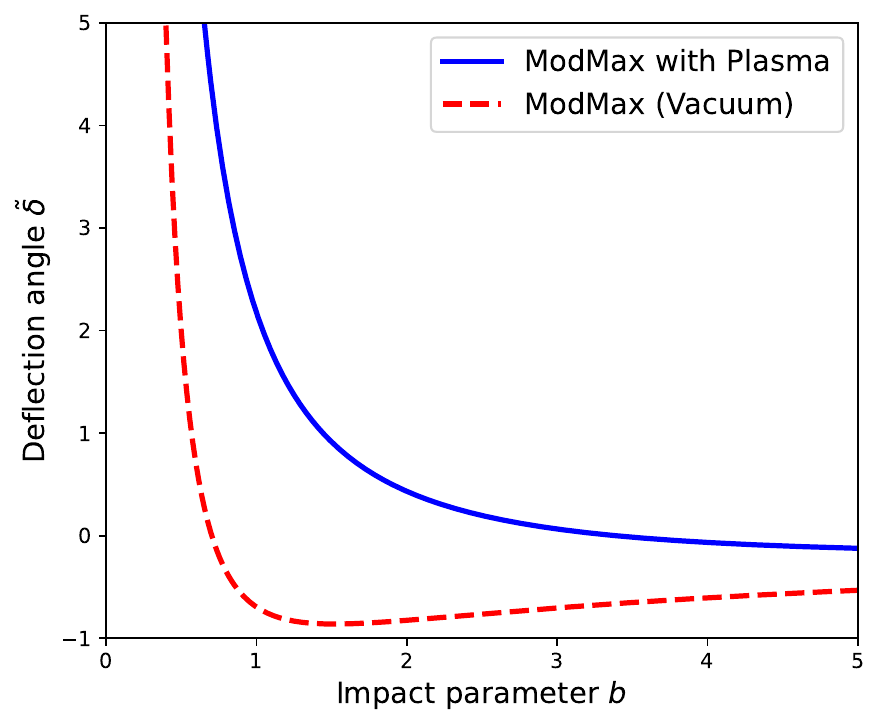}
\caption{Analysis of the deflection angle as a function of the impact parameter $b$ for the ModMax black hole with and without plasma. The plasma effect is characterized by $\omega_p^2 = 0.5$, while the vacuum case corresponds to $\omega_p^2 = 0$. The remaining parameters are fixed as $m_0 = 1$, $l = 10$, $q = 0.5$, $\gamma = 0.5$, and $\Lambda' = 0.45$.}

  \label{fig:da4}
\end{figure}

In the weak-field approximation, keeping linear contributions in
$m_0$ and $q^2 e^{-\gamma}$, the Gaussian curvature multiplied by the
optical surface element $dS=\sqrt{g^{\text{opt}}}\,dr\,d\phi$
becomes
\begin{equation}
\begin{aligned}
K\,dS =\;&
-\frac{m_0}{\Lambda'^{1/2} r^2}
\left(1 - \frac{\omega_e^2}{\omega_\infty^2}\right)
-\frac{2 q^2 e^{-\gamma}}{\Lambda'^{1/2} r^2}
\left(1 + \frac{\omega_e^2}{\omega_\infty^2}\right) \\
&-\frac{m_0^2}{2\Lambda'^{1/2} r^2}
\frac{\omega_e^2}{\omega_\infty^2}
-\frac{m_0}{\Lambda'^{3/2} r^4}
\left(\frac{3m_0}{2}-4 q^2 e^{-\gamma}\right) \\
&+\mathcal{O}(q^4, m_0^3)\,.
\end{aligned}
\end{equation}
To calculate the bending angle in the weak-field approximation using the Gauss–Bonnet theorem and to compare it with the non-plasma medium, we use the straight-line approximation.  $r=b/\sin\phi$.
The deflection angle can therefore be calculated using the expression~\cite{Ishihara:2016vdc}
\begin{equation}
\tilde{\delta}
=-\int_0^\pi\int_{b/\sin\phi}^{\infty} K\, dS .
\end{equation}
Performing the integration, we obtain the weak deflection angle of the
BTZ--ModMax black hole in plasma medium as
\begin{equation}
\begin{aligned}
\tilde{\delta}_{\text{ModMax}} =\;&
-\frac{2 m_0}{\Lambda'^{1/2} b}
-\frac{4 q^2 e^{-\gamma}}{\Lambda'^{1/2} b}
\ln\!\left(\frac{b}{l}\right)
\left(\frac{\omega_e^2}{\omega_\infty^2} + 1\right) \\
&+ \frac{2 m_0 \,\omega_e^2}{\Lambda'^{1/2} b \,\omega_\infty^2}
+ \frac{4 m_0 q^2 e^{-\gamma}}{9\, b^3 \Lambda'^{3/2}}
+ \mathcal{O}\!\left(\frac{m_0^2}{b^2},\,\frac{q^4}{b^2},\,\frac{1}{b^4}\right).
\end{aligned}
\end{equation}
From this expression, we see that the deflection angle in plasma depends
on the parameters $q$, $b$, $m_0$, $\Lambda'$ and the plasma frequency
ratio $\omega_e/\omega_\infty$. In Fig.~\ref{fig:da1}, we show the variation of the deflection angle as a function of the impact parameter $b$ for different values of $\Lambda'$. It is observed that the magnitude of the deflection angle decreases with increasing $\Lambda'$. This behavior arises due to the $1/\sqrt{\Lambda'}$ dependence, which effectively weakens the gravitational bending. For larger values of $b$, all curves tend to converge, indicating that the spacetime curvature effects become negligible in the weak-field regime.

In Fig.~\ref{fig:da2}, the deflection angle is plotted for different values of the ModMax parameter $\gamma$. As $\gamma$ increases, the deflection angle decreases for all values of $b$. This is due to the exponential suppression factor $e^{-\gamma}$, which reduces the contribution from the charge term. Physically, this shows that nonlinear electrodynamics effects tend to suppress the influence of charge, leading to weaker light bending. In Fig.~\ref{fig:da3}, we illustrate the effect of the plasma parameter $\omega_p^2$ on the deflection angle. It is clear that increasing $\omega_p^2$ enhances the deflection angle. This behavior is attributed to the refractive index of the plasma medium, which modifies photon trajectories and introduces frequency-dependent corrections. Even at large $b$, the separation between curves persists, indicating that plasma effects remain significant in the weak-field limit.

In Fig.~\ref{fig:da4}, we compare the deflection angle for the ModMax black hole in the presence and absence of plasma. The deflection angle is significantly enhanced in the plasma medium compared to the vacuum case. The difference is more pronounced at smaller values of $b$, where both gravitational and dispersive effects are stronger. This demonstrates that plasma effects play a crucial role and can dominate over nonlinear electrodynamics corrections in determining the bending of light. In summary, these results highlight the interplay between spacetime curvature, nonlinear electrodynamics, and plasma effects, showing that the presence of a dispersive medium can significantly modify the gravitational lensing behavior around BTZ--ModMax black holes.


\section{Bound on Greybody Factor of BTZ--ModMax Black Hole in Plasma Medium} \label{sec5}

Greybody factors (transmission probability) characterize the deviation from a perfect blackbody spectrum due to spacetime curvature and matter fields surrounding the black hole. In this section, we derive the rigorous bound on the greybody factor for a BTZ black hole coupled to ModMax nonlinear electrodynamics in the presence of a cold, non-magnetized plasma.

The general bound on the transmission probability is given by~\cite{Konoplya:2011qq,Boonserm:2023oyt}
\begin{equation}
T \ge \text{sech}^2 \left( \frac{1}{2\omega} \int_{-\infty}^{\infty} V_{\text{s}}(r)\, dr_* \right),
\end{equation}
where $r_*$ is the tortoise coordinate defined as
\begin{equation}
dr_* = \frac{dr}{f(r)},
\end{equation}
and $\omega$ is the frequency of the scalar perturbation.

For the BTZ--ModMax black hole, the metric function is given by
\begin{equation}
f(r)=\Lambda' r^2 - m_0 - 2 q^2 e^{-\gamma}\ln\!\left(\frac{r}{l}\right).
\end{equation}

In the presence of plasma, the refractive index modifies the scalar potential. The corresponding Regge--Wheeler equation takes the form\cite{regge1957stability}
\begin{equation}
\left(\frac{d^2}{dr_*^2} + \omega^2 - V_{\text{s}}(r)\right)\psi = 0,
\end{equation}
where the scalar potential in $(2+1)$ dimensions becomes
\begin{equation}
V_{\text{s}}(r)
=
f(r)\left[
\frac{\ell^2}{r^2}
+ \frac{f'(r)}{2r}
- \frac{f(r)}{4r^2}
\right]
+ f(r)\frac{\omega_e^2}{\omega_\infty^2}.
\end{equation}

Using the tortoise coordinate, the bound can be rewritten as
\begin{equation}
T \ge \text{sech}^2 \left( \frac{1}{2\omega} \int_{r_+}^{\infty} \frac{V_{\text{s}}(r)}{f(r)}\, dr \right),
\end{equation}
where $r_+$ denotes the event horizon determined from $f(r_+)=0$.

Substituting the scalar potential, we obtain
\begin{equation}
\frac{V_{\text{s}}(r)}{f(r)}
=
\frac{\ell^2}{r^2}
+ \frac{f'(r)}{2r}
- \frac{f(r)}{4r^2}
+ \frac{\omega_e^2}{\omega_\infty^2}.
\end{equation}

Using
\begin{equation}
f'(r)=2\Lambda' r - \frac{2 q^2 e^{-\gamma}}{r},
\end{equation}
the above expression simplifies to
\begin{equation}
\begin{aligned}
\frac{V_{\text{s}}(r)}{f(r)}
=&\;
\frac{\ell^2}{r^2}
+ \Lambda'
- \frac{q^2 e^{-\gamma}}{r^2}
- \frac{\Lambda'}{4}
+ \frac{m_0}{4r^2} \\
&+ \frac{q^2 e^{-\gamma}}{2r^2}\ln\!\left(\frac{r}{l}\right)
+ \frac{\omega_e^2}{\omega_\infty^2}.
\end{aligned}
\end{equation}

The integral appearing in the bound diverges due to the asymptotic non-flatness of the BTZ spacetime and the constant plasma contribution. Therefore, we regularize the integral by subtracting the asymptotic contribution, yielding a finite expression.

Performing the integration from $r_+$ to $\infty$, we obtain
\begin{equation}
\begin{aligned}
\int_{r_+}^{\infty} \frac{V_{\text{s}}(r)}{f(r)}\, dr
=&\;
\frac{\ell^2}{r_+}
+ \frac{m_0}{4r_+}
- \frac{q^2 e^{-\gamma}}{r_+} \\
&+ \frac{q^2 e^{-\gamma}}{2r_+}
\left[
\ln\!\left(\frac{r_+}{l}\right) + 1
\right].
\end{aligned}
\end{equation}

Thus, the rigorous bound on the greybody factor for the BTZ--ModMax black hole in plasma medium is given by
\begin{equation}
T \ge \text{sech}^2 \left[
\frac{1}{2\omega}
\left(
\frac{\ell^2}{r_+}
+ \frac{m_0}{4r_+}
- \frac{q^2 e^{-\gamma}}{r_+}
+ \frac{q^2 e^{-\gamma}}{2r_+}
\left(\ln\!\frac{r_+}{l} + 1\right)
\right)
\right].
\end{equation}

\begin{figure}[ht]
  \centering
  \includegraphics[width=85mm,height=75mm]{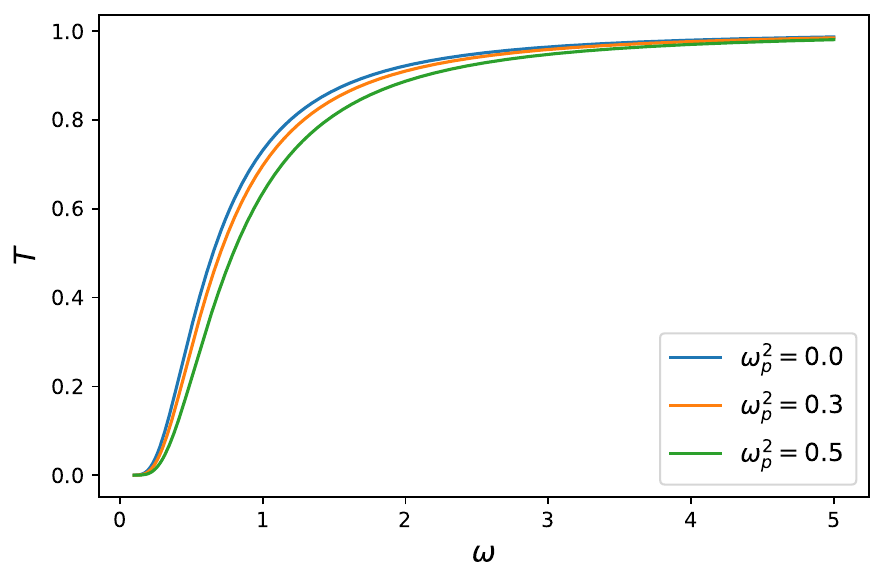}
\caption{Greybody factor bound $T$ versus frequency $\omega$ for different values of the plasma parameter $\omega_{p}^{2}$.  The remaining parameters are fixed at $m_{0}=0.8$, $\gamma=0.2$, $q=0.5$,$\Lambda=1$, and $l=1$.}

  \label{fig:gb1}
\end{figure}

From the above expression, it is evident that the greybody factor depends on the black hole parameters $m_0$, $q$, $\gamma$, $\Lambda'$, and the angular momentum $\ell$, as well as the plasma frequency ratio $\omega_e/\omega_\infty$. 
In Fig.~\ref{fig:gb1}, we present the behavior of the greybody factor bound $T$ as a function of frequency $\omega$ for different values of the homogeneous plasma parameter $\omega_p^2$. It is evident that increasing $\omega_p^2$ leads to a suppression of the transmission probability in the low-frequency regime. 
This behavior can be attributed to the dispersive nature of the homogeneous plasma, which effectively increases the refractive index and introduces a stronger potential barrier for low-energy modes. As a result, the propagation of low-frequency radiation is hindered. 
However, in the high-frequency limit, all curves converge to $T \to 1$, indicating that high-energy modes are minimally affected by the plasma environment. 
Thus, the influence of the homogeneous plasma is predominantly significant in the low-energy sector of the spectrum.

\begin{figure}[ht]
  \centering
  \includegraphics[width=85mm,height=75mm]{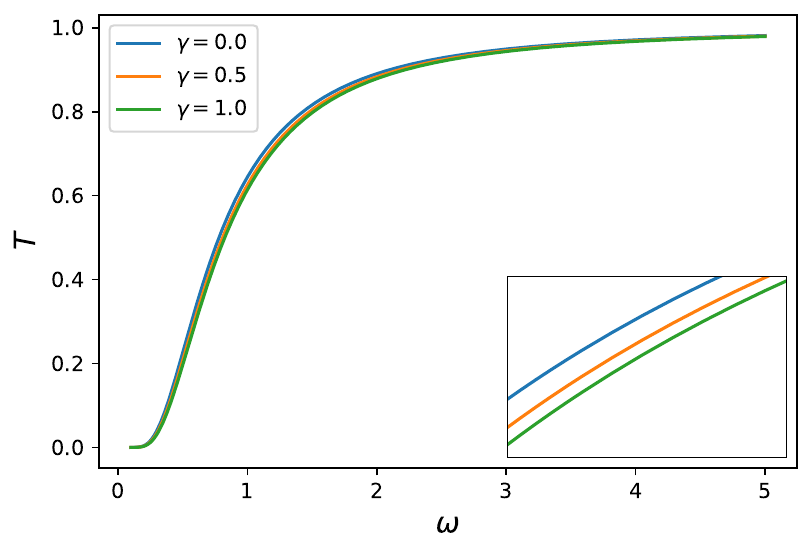}
\caption{Greybody factor bound $T$ versus frequency $\omega$ for different values of the ModMax parameter $\gamma$. The remaining parameters are fixed at $m_{0}=0.8$, $\omega_{p}^{2}=0.5$, $q=0.5$,$\Lambda=1$, and $l=1$. The inset displays an enlarged perspective of the area.  $1.2 \leq \omega \leq 1.5$ and $0.7 \leq T \leq 0.8$, highlighting the differences among the curves.}

  \label{fig:gb2}
\end{figure}

In Fig.~\ref{fig:gb2}, we illustrate the variation of the greybody factor bound $T$ with frequency $\omega$ for different values of the ModMax plasma parameter $\gamma$. 
We observe that an increase in $\gamma$ slightly reduces the transmission probability, particularly in the intermediate frequency range. This indicates that the nonlinear electrodynamics effects associated with the ModMax parameter modify the effective potential experienced by the propagating waves. The inset clearly highlights these small but systematic deviations among the curves.  In contrast, the high-frequency regime remains largely unaffected, suggesting that nonlinear corrections become negligible for high-energy modes. Physically, this implies that the ModMax parameter introduces subtle corrections to wave propagation in the plasma surrounding the black hole, which may have implications for the spectrum of emitted radiation.

\section{Conclusion} \label{sec6}

The present research examines the deflection of light and greybody factor in a homogeneous plasma medium around a BTZ--ModMax black hole. Although it is well known that $(2+1)$-dimensional black hole geometries are not directly astrophysical, they provide a useful theoretical framework to explore fundamental aspects of gravity and nonlinear electrodynamics. In this context, our analysis offers a mathematical framework to study the effects of nonlinear electrodynamics in the presence of an astrophysical-like environment, such as a plasma medium. The main results can be summarized as follows:

\begin{itemize}

\item The photon orbit radius $r_{\mathrm{ph}}$ decreases with increasing charge $q$ and increases with $\gamma$. This indicates that the charge induces an effective inward shift of the photon sphere, while the nonlinear electrodynamics parameter $\gamma$ produces a mild outward correction.

\item The presence of plasma modifies photon motion through the plasma frequency term $\omega_p^2$, leading to a frequency-dependent refractive index. This generally shifts the photon sphere outward and introduces deviations from the vacuum case.

\item The deflection angle decreases with increasing $\Lambda'$, , indicating a reduction in gravitational bending  due to the effective cosmological contribution.

\item An elevation in the ModMax parameter $\gamma$ lowers te deflection angle, as the exponential factor $e^{-\gamma}$ suppresses the influence of the charge term.

\item The plasma parameter $\omega_p^2$ enhances the deflection angle, demonstrating that dispersive effects play a significant role in modifying photon trajectories even in the weak-field regime.

\item A direct comparison between plasma and vacuum cases shows that plasma significantly enhances light bending, especially at small impact parameters, highlighting its dominant role over nonlinear electrodynamics corrections.

\item The greybody factor analysis reveals that increasing $\omega_p^2$ suppresses the transmission probability in the low-frequency regime due to the enhanced effective potential barrier introduced by the plasma.

\item The ModMax parameter $\gamma$ introduces subtle corrections to the greybody factor, slightly reducing transmission probabilities in the intermediate frequency range, while high-frequency modes remain largely unaffected.

\item In the high-frequency limit, the greybody factor approaches unity ($T \to 1$) for all cases, indicating that high-energy radiation is insensitive to both plasma and nonlinear electrodynamics effects.

\end{itemize}

In conclusion, our results demonstrate that the combined effects of spacetime curvature, nonlinear electrodynamics, and plasma environment play a crucial role in shaping light propagation and wave emission processes around BTZ--ModMax black holes, with potential implications for gravitational lensing and observational signatures. As a future direction, it would be interesting to extend this analysis to $(3+1)$-dimensional ModMax black hole spacetimes in the presence of plasma, in order to investigate the optical properties and assess their relevance in more realistic astrophysical scenarios.

\section*{Acknowledgements}

The author, SK, sincerely acknowledges IMSc for providing exceptional research facilities and a conducive environment that facilitated his work as an Institute Postdoctoral Fellow. The author H.N. acknowledges financial support from the Anusandhan National Research Foundation (ANRF), through the Science and Engineering Research Board (SERB) Core Research Grant (Grant No. CRG/2023/008980).

\bibliographystyle{apsrev4-2}
\bibliography{modmaxplasma}

\end{document}